\documentclass{ws-ijmpcs}

\begin{document}

\markboth{W. Zhao \& L. Santos} {The weird side of the Universe:
Preferred axis}

%
\catchline{}{}{}{}{}
%

\title{The weird side of the Universe: Preferred axis}

\author{Wen Zhao\footnote{wzhao7@ustc.edu.cn} and Larissa Santos}

\address{Department of Astronomy, University of Science and Technology of China, \\
Hefei, Anhui 230026, P.R.China}



\maketitle

\begin{history}
\received{Day Month Year}
\revised{Day Month Year}
\published{Day Month Year}
\end{history}

\begin{abstract}
{In both WMAP and Planck observations on the temperature
anisotropy of cosmic microwave background (CMB) radiation a
number of large-scale anomalies were discovered in the past years, including the CMB
parity asymmetry in the low multipoles. By defining a
directional statistics, we find that the CMB parity asymmetry is
directional dependent, and the preferred axis is stable, which means that it is
independent of the chosen CMB map, the definition of the statistic, or
the CMB masks. Meanwhile, we find that this preferred axis strongly
aligns with those of the CMB quadrupole, octopole, as well as those of
other large-scale observations. In addition, all of them aligns
with the CMB kinematic dipole, which hints to the non-cosmological
origin of these directional anomalies in cosmological
observations.}
\end{abstract}

\ccode{PACS numbers: 95.85.Sz, 98.70.Vc, 98.80.Cq}

\section{Introduction}
In the past twenty years, based on various cosmological
observation, including the temperature and polarization
anisotropies of the cosmic microwave background (CMB) radiation, the
distribution of the galaxies, Type Ia supernovas, the weak
lensing, and so on, the so-called standard
cosmological model, i.e. inflation+$\Lambda$CDM model
\cite{planck-cosmology}, has been built. In this model, the universe is completely
described by six parameters, i.e. the energy density of baryon
$\Omega_b$ and dark matter $\Omega_{\rm CDM}$, the Hubble constant
$H_0$, the optical depth of reionization $\tau$, the amplitude
of primordial density perturbations $A_s$, and the spectral index
$n_s$. This successful model is based on the following
assumptions: (1) On large scales, the Universe is isotropic and
homogeneous, known as the cosmological principle;
(2) Einstein's General Relativity is the correct theory that
describes gravity on all the macroscopic scales; (3) The main
components of the Universe are baryons, cold dark matter and dark
energy (or cosmological constant $\Lambda$); (4) Primordial
fluctuations were created as quantum fluctuations, which gave rise
to structure formation.

At the same time, with the release of various precise observed
data, a number of large-scale ``anomalies" have also been reported
recently. In particular, it was noticed that some of them are
directional dependent, e.g. the alignment of CMB low multipoles,
the large-scale velocity flows, the alignment of the polarization of QSOs,
the directional dependence of CMB parity asymmetry, the anisotropy of
cosmic acceleration, the anisotropy of the fine structure constant
$\alpha$ and so on \cite{axis-universe}. If this kind of
directional anomaly has a cosmological origin, they will
challenge the standard cosmological model, and change the base of
modern cosmology. In this paper, we shall first focus on the
directional dependence of the CMB parity asymmetry by searching
for the preferred axis stored in it, and compare this axis with
the other ones in the other observations. In addition, we shall
also introduce the possible physical origins of these anomalies.




\section{The parity asymmetry in the CMB temperature anisotropy}
By the observations of WMAP and Planck satellites, people found
that in the high multipole range, i.e. in the small scales,
the observed data excellently fit the theoretical prediction.
However, in the low multipole range $\ell<100$, the data are quite
problematic. Some non-gaussian anomalies are reported in
this scale, which includes the low quadrupole problem, the lack of
large-scale correlation, the cold spot, the CMB power asymmetry,
the mirror asymmetry, the large-scale quadrant asymmetry, the alignment of low
multipoles, the parity asymmetry and so on
\cite{wmap-anomalies,planck2013-anomalies}. Based on these facts,
the Planck collaboration claimed that ``the Universe is still
weird and interesting". Here, let us focus on the CMB parity
asymmetry.

We can decompose the two-dimensional CMB temperature anisotropy field in the
sphere as follows,
\begin{equation}
\Delta T(\theta,\phi)=\sum_{\ell=0}^{\infty}\sum_{m=-\ell}^{\ell}
a_{\ell m} Y_{\ell m}(\theta,\phi),
\end{equation}
where $a_{\ell m}$ are coefficients which satisfy the Gaussian
distribution in the standard inflationary scenario. The power
spectrum is defined as $C_{\ell}=\langle a_{\ell m} a^*_{\ell m}
\rangle$, where the bracket denotes the ensemble average. In order
to estimate the power spectrum, one can define the unbiased
estimator as
$\hat{C}_{\ell}=\frac{1}{2\ell+1}\sum_{m=-\ell}^{\ell} a_{\ell
m}a^*_{\ell m}$, which satisfies the $\chi^2$-distribution with
the expectation value $\langle \hat{C}_{\ell}\rangle=C_{\ell}$.
This means that, in principle, the data $\hat{C}_{\ell}$ should
randomly oscillate around theoretical power spectra $C_{\ell}$.

However, in the real data, it was noticed that in the
low-multipole range $\ell\lesssim 30$, the even-multipole data are
systematical smaller than the theoretical curve, while the odd
ones are systematically larger than the model predictions, which
is the so-called CMB parity asymmetry. By defining the statistic
as the ratio between the sum of all the even multipoles and
that of the odd ones, people found that the anomalies maximize at
about $\ell_{\max}=22$ \cite{kim-review}.


\section{Directional properties of CMB parity violation}

In order to study the direction properties of the CMB parity
asymmetry, we define the new unbiased estimator for $C_{\ell}$ as
follows \cite{zhao-2012},
\begin{equation}
D_{\ell}=\frac{1}{2\ell} \sum_{m=-\ell}^{\ell} a_{\ell m}
a^*_{\ell m}(1-\delta_{m0}).
\end{equation}
Comparing with the standard one, $\hat{C}_{\ell}$, this new
estimator is rotationally variant, its value depends on the
choice of the coordinate system, and the preferred axis is exactly
the $z$-axis of the coordinate system. For any given coordinate system
with the $z$-direction labelled as $\hat{\rm q}$ (which can also
be treated as the coordinate of this direction in the Galactic
coordinate system), we denote the corresponding estimator as
$D_{\ell}(\hat{\rm q})$.

Now, we can define the rotationally variable parity parameter
$G_1(\ell;\hat{\rm q})$ by using $D_{\ell}(\hat{\rm q})$ as
follows:
\begin{equation}
G_1(\ell;\hat{\rm q})\equiv
\frac{\sum_{\ell'=2}^{\ell_{\max}}{(2\ell'+1)}D_{\ell'}(\hat{\rm
q})\Gamma^{+}(\ell')}{\sum_{\ell'=2}^{\ell_{\max}}{(2\ell'+1)}D_{\ell'}(\hat{\rm
q})\Gamma^{-}(\ell')}
\end{equation}
where $\Gamma^{+}(\ell)=\cos^2(\ell\pi/2)$ and
$\Gamma^{-}(\ell)=\sin^2(\ell\pi/2)$. By considering all the
possible $\hat{\rm q}$, we can construct the 2-dimensional
$G_1$-map for any given maximum multipole $\ell$, in which the
smaller $G_1$ value denotes the larger parity asymmetry. In this
paper, we define the direction in which the $G_1$ value is
minimized as the preferred axis.

In order to search for the preferred axis in the CMB parity
asymmetry, we consider the WMAP ILC7 map, and construct the
corresponding $G_1$-map for each maximum $\ell$. We find that for
any given $\ell$, the morphologies of these $G_1$-maps are similar
to each other, and their preferred axes are all around
($\theta=45^{\circ}$, $\phi=280^{\circ}$)\cite{zhao-2012}. The
angles between the axes of different maximum are smaller than
$15^{\circ}$ as long as $3<\ell<22$.


To study the stability of our conclusion, we apply the directional
analysis to the Planck observations, including the Commander,
NILC, SMICA and SEVEM maps. From the results of Commander, NILC
and SMICA maps, we find quite similar results, which are all
consistent with the ones derived from the WMAP ILC7 map\cite{zhao-2014}. The results of SEVEM map are quite different. This is caused by the extremely dirty region of the Galactic plane region on this map.


The results derived above are all based on the definition of
$G_1$ statistic. However, an important problem arises: Whether or
not the preferred axis of CMB parity asymmetry depends on the
definition of statistic or estimator? In order to cross-check the
results, we define the following directional statistics $G_i$
($i=2,3,4,5,6$) \cite{zhao-2014}, which are quite different from
$G_1$.
\begin{equation}
G_2(\ell;\hat{\rm q})\equiv
\frac{\sum_{\ell'=2}^{\ell_{\max}}{\ell'(\ell'+1)}D_{\ell'}(\hat{\rm
q})\Gamma^{+}(\ell')}{\sum_{\ell'=2}^{\ell_{\max}}{\ell'(\ell'+1)}D_{\ell'}(\hat{\rm
q})\Gamma^{-}(\ell')},~ G_3(\ell;\hat{\rm
q})\equiv\frac{2}{\ell-1}\sum_{\ell'=3}^{\ell}
\frac{(\ell'-1)\ell'D_{\ell'-1}(\hat{\rm
q})}{\ell'(\ell'+1)D_{\ell'}(\hat{\rm q})}
\end{equation}
$G_{i}$ ($i=4,5,6$) are same with $G_{i}$ ($i=1,2,3$) but the
estimators $D_{\ell}$ are replaced by
$\tilde{D}_{\ell}=\frac{1}{2\ell+1}\sum_{m} m^2|a_{\ell m}|^2$.
Then, we repeat our analysis by adopting the new statistics.
Interesting enough, we find that the morphologies of the
$G_i$-maps are completely different for different statistics.
However, their preferred axes are nearly same for all the
used statistics and all the maximum multipole $\ell$, as long as
$\ell<10$ \cite{zhao-2014}. So, we conclude that the preferred
axis of the CMB parity asymmetry is independent of the definition of
the directional statistic.

In the CMB observations, various foreground residuals are always
unavoidable, especially in the Galactic region. It is worthy to
investigate the cases in which these contaminated data are
excluded. The simplest way to exclude the polluted region is to
apply the top-hat mask to the data. In order to study the effect
of CMB mask on the preferred axis. We consider the Planck data and
taking into account the corresponding mask suggested by Planck
team. For each CMB map and the mask, we define the
pseudo-estimator of CMB power spectra as follows \cite{zhao-2016},
$\tilde{D}_{\ell}=\frac{1}{2\ell} \sum_{m=-\ell}^{\ell}
\tilde{a}_{\ell m}\tilde{a}_{\ell m}^*(1-\delta_{m0})$,
where $\tilde{a}_{\ell m}$ are the pseudo-coefficients of the
masked CMB map. The unbiased estimator is defined as
$\mathcal{D}_{\ell}=\sum_{\ell'} N_{\ell \ell'}^{-1}
\tilde{D}_{\ell'}$,
where
\begin{eqnarray}
N_{\ell \ell'}&=&M_{\ell \ell'}-\frac{2\ell' +1}{2\ell}
\sum_{\ell_{2} \ell_{2}^{'}m_{1}} \frac {\sqrt{(2\ell_{2}
+1)(2\ell_{2}^{'} +1)}} {4\pi} \times \nonumber \\
&&
\begin{pmatrix}
    \ell'&\ell_2&\ell \\
    0 & 0& 0
\end{pmatrix}
\begin{pmatrix}
    \ell'&\ell_2^{'}&\ell \\
    0 & 0& 0
\end{pmatrix}
\begin{pmatrix}
    \ell'&\ell_2&\ell \\
   m_1 & -m_1& 0
\end{pmatrix}
\begin{pmatrix}
    \ell'&\ell_2&\ell \\
    m_1 & -m_1& 0
\end{pmatrix} w_{\ell_2 m_1} w_{\ell_2^{'} m_1},
\end{eqnarray}
and
\begin{equation}
M_{\ell \ell'}=(2\ell'+1)\sum_{\ell_{2}} \frac {2\ell_{2}+1}
{4\pi}
\begin{pmatrix}
    \ell'&\ell_2&\ell \\
    0 & 0& 0
    \end{pmatrix}  ^{2}\tilde{w}_{\ell_{2}}.
\end{equation}
We define the directional statistic by a similar way as
$G_1(\ell;\hat{\rm q})$, but the estimator $D_{\ell}$ is replaced
by $\mathcal{D}_{\ell}$. Applying the described method to the masked
Planck maps, we find nearly the same preferred axis as that in the
full-sky map \cite{zhao-2016}.

\section{Comparison with the other direction-dependent anomalies}

As well known, the lowest anisotropy of CMB is the dipole
component with an amplitude of 3.35 mK, and it is caused by the peculiar
velocity of the Solar System relative to the comoving cosmic rest
frame \cite{dipole}.  Relative to the observers, the dipole
anisotropy defines a peculiar axis in the Universe, which is at
($\theta=42^{\circ}$, $\phi=264^{\circ}$) in the Galactic
coordinate system. By comparing it with the preferred axis
discovered above, we find the strong alignment between them. The
angle between them is smaller than $10^{\circ}$.

The lowest cosmological anisotropic modes of the CMB fluctuations
are the quadrupole and octopole. By defining the proper
directional statistics, people discovered their preferred axes \cite{quadrupole} (see Table 1). In addition, they are strongly correlated, and very close to the
direction defined by the CMB kinematic dipole. In order to study
the relation between these axes and the one discovered here, we define
the average angle between these four axes (CMB dipole, quadrupole,
octopole, and the CMB parity asymmetry). Comparing with the random
simulations, we find that the alignment between them are confirmed
at more than $3\sigma$ \cite{zhao-2014}.

Besides the peculiar axes in the CMB, several other preferred axes
are also reported in various large-scale cosmological
observations, including the alignment of quasar polarization
vectors, large-scale velocity flows of the cosmic matter in the
CMB rest frame, the distribution of handedness of spiral galaxies,
the anisotropy of the cosmic acceleration, the anisotropic
distribution of fine-structure constant \cite{axis-universe}. We
list the direction of these axes in Table 1, in which we find that
all the large-scale observations point to the nearly same
preferred direction. And also, this direction is exactly the same
direction defined by CMB kinematic dipole. So, it is also called
the \emph{evil axis} in cosmology.

\begin{table} 
\tbl{Preferred axis in various large-scale observations.}
{\begin{tabular}{@{}cccc@{}}
\toprule observations & $\theta$ [degree] & $\phi$ [degree] \\
\colrule
CMB parity asymmetry & 45.82 & 279.73  \\
CMB kinematic dipole & 42 & 264  \\
CMB quadrupole & 13.4 & 238.5  \\
CMB octopole & 25.7 & 239.0  \\
Polarization of QSOs & 69 & 267  \\
Large-scale velocity flows & 84 & 282  \\
Handedness of spiral galaxies & 158.5 & 232  \\
Anisotropy of cosmic acceleration & 23.4 & 247.5  \\
Distribution of fine-structure constant & 104 & 331
\\ \botrule
\end{tabular} \label{ta1}}
\end{table}

\section{Possible explanations}

The standard cosmological model is based on two assumptions: One
is that Einstein's general relativity correctly describes gravity,
the other assumes the universe as homogeneous and isotropic on
large scales. If we believe that the anomalies have a cosmological
origin, at least one of these two assumptions will be broken
\cite{axis-universe}. One possibility relies on the Bianchi
models. The Bianchi classification provides a complete
characterization of all the known homogeneous but anisotropic
exact solution to general relativity. So, in general, Bianchi
models can provide preferred directions in the universe. Another
way is to revise the gravitation theory. For instance, some
authors considered that the universe is influenced by large-scale
¡°wind¡±, and the cosmic matter is drifted by this ¡°wind¡±, which
is described by the Finsler geometry \cite{axis-universe}.


On the contrary, some other people believe that these anomalies
are due to some non-cosmological reasons: Unsolved systematical
errors, calibration errors or foreground contaminations (CMB
dipole-related). One possible reason is related to the contaminations
generated by the collective emission of Kuiper Belt objects.
Another explanation may relate it to a deviation measured in the CMB
kinematic dipole. It is also possible that the preferred direction
is caused by the tidal field originated from the anisotropy of our
local halo \cite{zhao-2012,axis-universe}.

\section{Discussions and conclusions}

In the recent observations of the large-scale structure, several
directional anomalies have been reported, including anomalies in
the CMB low multipoles, and the CMB parity asymmetry. Although the
confidence level for each individual anomaly is not too high, the
directional alignment of all these anomalies is quite significant,
which strongly suggests a common origin of these anomalies.

If these anomalies are due to cosmological effects, e.g. an
alternative theory of gravity or geometry, they indicate the
violation of the cosmological principle. So, one should consider building a new cosmological model to explain the large-scale data.
However, if these directional anomalies arise from
non-cosmological reasons, e.g. the unsolved systematical errors or
contaminations, we should carefully treat the current data, and
exclude the errors in the future analysis to avoid the misleading
explanations of the data. Although the physical origins are still
unclear, from the alignment between preferred axes of cosmological
observations and the motion of the Solar System in the CMB rest frame,
we are lead to believe the non-cosmological origin of the
large-scale anomalies. We expect that the future measurements on
the CMB polarization fields, the cosmic weak lensing, or the
distribution of 21-cm line can help us to solve the puzzles.

\section*{Acknowledgments}

This work is supported by NSFC No. 11603020, 11633001, 11173021, 11322324, 11653002, 11421303, project of Knowledge Innovation Program of Chinese Academy of Science, the Fundamental Research Funds for the Central Universities and the Strategic Priority Research Program of the Chinese Academy of Sciences Grant No.XDB23010200.



\begin{thebibliography}{0}    

\bibitem{planck-cosmology} Planck Collab. (P. A. R. Ade {\it et al}.), {\it Astron. Astrophys.} {\bf 571}, A1 (2014).

\bibitem{axis-universe} see for instance, W. Zhao and L. Santos, {\it The
Universe} {\bf 3}, 9 (2015).

\bibitem{wmap-anomalies}
WMAP Collab. (C. L. Bennett {\it et al.}), {\it Astrophys. J.
Suppl.} {\bf 192}, 17 (2011).

\bibitem{planck2013-anomalies}
Planck Collab. (P. A. R. Ade {\it et al}.), {\it Advances in
Astronmy} {\bf 571}, A23 (2014).


\bibitem{kim-review} J. Kim, P. Naselsky and M. Hansen, {\it Advances in
Astronmy} {\bf 2012}, 960509 (2012).

\bibitem{zhao-2012} P. Naselsky, W. Zhao, J. Kim and S. Chen, {\it Astrophys. J.} {\bf 748}, 31 (2012).

\bibitem{zhao-2014} W. Zhao, {\it Phys. Rev. D} {\bf 89}, 023010
(2014).

\bibitem{zhao-2016} C. Cheng, W. Zhao, Q. Huang and L. Santos, {\it Phys. Lett.
B} {\bf 757}, 445 (2016).

\bibitem{dipole}
WMAP Collab. (N. Jarosik {\it et al.}), {\it Astrophys. J. Suppl.}
{\bf 192}, 14 (2011).

\bibitem{quadrupole}
M. Tegmark, A. de Oliveira-Costa and A. Hamilton, {\it Phys. Rev.
D} {\bf 68}, 123523 (2003).


\end{thebibliography}
\end{document}